\documentclass[reprint,superscriptaddress,groupedaddress,amsmath,amssymb,aps,pre]{revtex4}
\usepackage{graphicx}  
\usepackage{dcolumn}   
\usepackage{bm}        
\usepackage{tabularx}
\usepackage{array}
\usepackage[LGR,T1]{fontenc}
\usepackage[latin9]{inputenc}
\usepackage{textcomp}
\usepackage{amstext}

\usepackage{float}
\usepackage[caption = false]{subfig}

\usepackage{color} 
\usepackage{ulem} 
\usepackage{hyperref}

\begin{document}

\title{High-Sensitivity Temperature Sensing Using an Implanted Single Nitrogen-Vacancy
Center Array in Diamond}

\author{Junfeng Wang}
\affiliation{Hefei National Laboratory for Physical Science at Microscale, and
Department of Physics, University of Science and Technology of China,
Hefei, Anhui, 230026, P. R. China}
\author{ Fupan Feng}
\affiliation{Hefei National Laboratory for Physical Science at Microscale, and
Department of Physics, University of Science and Technology of China,
Hefei, Anhui, 230026, P. R. China}
\author{Jian Zhang}
\affiliation{Hefei National Laboratory for Physical Science at Microscale, and
Department of Physics, University of Science and Technology of China,
Hefei, Anhui, 230026, P. R. China}
\author{Jihong Chen}
\affiliation{Accelerator Laboratory, School of Physics and Technology, Wuhan University,
Wuhan, Hubei, 430072, P. R. China}
\author{Zhongcheng Zheng}
\affiliation{Accelerator Laboratory, School of Physics and Technology, Wuhan University,
Wuhan, Hubei, 430072, P. R. China}
\author{Liping, Guo}
\affiliation{Accelerator Laboratory, School of Physics and Technology, Wuhan University,
Wuhan, Hubei, 430072, P. R. China}
\author{Wenlong Zhang}
\affiliation{Hefei National Laboratory for Physical Science at Microscale, and
Department of Physics, University of Science and Technology of China,
Hefei, Anhui, 230026, P. R. China}
\author{Xuerui Song}
\affiliation{Hefei National Laboratory for Physical Science at Microscale, and
Department of Physics, University of Science and Technology of China,
Hefei, Anhui, 230026, P. R. China}
\author{Guoping Guo}
\affiliation{Key lab of Quantum Information, CAS, University of Science and Technology
of China, Hefei, Anhui, 230026, P. R. China}
\author{Lele Fan}
\affiliation{National Synchrotron Radiation Laboratory, University of Science
and Technology of China, Hefei, 230029, P. R. China}
\author{Chongwen Zou}
\affiliation{National Synchrotron Radiation Laboratory, University of Science
and Technology of China, Hefei, 230029, P. R. China}
\author{Liren Lou}
\affiliation{Hefei National Laboratory for Physical Science at Microscale, and
Department of Physics, University of Science and Technology of China,
Hefei, Anhui, 230026, P. R. China}
\author{Wei Zhu}
\affiliation{Hefei National Laboratory for Physical Science at Microscale, and
Department of Physics, University of Science and Technology of China,
Hefei, Anhui, 230026, P. R. China}
\author{Guanzhong Wang}\email{gzwang@ustc.edu.cn}
\affiliation{Hefei National Laboratory for Physical Science at Microscale, and
Department of Physics, University of Science and Technology of China,
Hefei, Anhui, 230026, P. R. China}

\begin{abstract}
We presented a high-sensitivity temperature detection using an implanted single Nitrogen-Vacancy center array in diamond. The high-order Thermal Carr-Purcell-Meiboom-Gill (TCPMG)
method was performed on the implanted single nitrogen vacancy (NV)
center in diamond in a static magnetic field. We demonstrated that
under small detunings for the two driving microwave frequencies, the
oscillation frequency of the induced fluorescence of the NV center
equals approximately to the average of the detunings of the two driving
fields. On basis of the conclusion, the zero-field splitting D for
the NV center and the corresponding temperature could be determined.
The experiment showed that the coherence time for the high-order TCPMG
was effectively extended, particularly up to 108 $\mu s$ for
TCPMG-8, about 14 times of the value 7.7 $\mu s$ for thermal
Ramsey method. This coherence time corresponded to a thermal sensitivity
of 10.1 mK/Hz$^{1/2}$. We also detected the temperature
distribution on the surface of a diamond chip in three different circumstances
by using the implanted NV center array with the TCPMG-3 method. The
experiment implies the feasibility for using implanted NV centers
in high-quality diamonds to detect temperatures in biology, chemistry,
material science and microelectronic system with high-sensitivity
and nanoscale resolution.
\end{abstract}

\maketitle

In recent years some thermal detection techniques have been developed
to map temperature distribution with spatial resolution down to micrometer-nanometer
range\cite{key-1}, such as Raman spectroscopy\cite{key-1,key-2},
fluorescence thermography\cite{key-1,key-3}, and scanning thermal
microscopy.\cite{key-4} However, such techniques are reported with
limitations like low sensitivity\cite{key-1,key-2} and large random
errors come from fluorescence rate fluctuations or fluorescence blinking
and bleaching in the local environment.\cite{key-1,key-2,key-3} Recently
the negatively charged nitrogen vacancy (NV$^-$)
center in diamond \cite{key-5,key-6,key-7,key-8,key-9,key-10} and
the spin defects in silicon carbide \cite{key-11} are investigated
as promising nanoscale temperature sensors with both high temperature
precision and high spatial resolution.\cite{key-8,key-9,key-10} 

The NV center is a spin defect consisting of a substitutional nitrogen
impurity adjacent to a carbon vacancy in diamond. It has increasingly
attracted attention in recent years owing to its excellent properties,
like photostability, biocompatibility, chemical inertness, and long
spin coherence and relaxation times ($\sim$ms in the isotopically
pure diamond) at room temperature. These remarkable properties have
been explored in many applications like quantum information processing,\cite{key-12,key-13,key-14,key-15,key-16}
metrologies such as magnetic field sensing,\cite{key-17,key-18,key-19}
electric field sensing,\cite{key-20,key-21} force sensing,\cite{key-22,key-23}
thermal sensing,\cite{key-8,key-9,key-10} single electron and nuclear
spin sensing,\cite{key-24,key-25,key-26} and external nuclear spin
sensing.\cite{key-27,key-28} In thermal sensing, Neumann et al. demonstrated
the measurement of the temperature distribution on a glass coverslip
using single NV center nanodiamonds as temperature sensors.\cite{key-9}
However, the thermal sensitivity was unsatisfactory due to the short
coherence time. To address the short coherence time issue, Toyli et
al. proposed the thermal Carr-Purcell-Meiboom-Gill
(TCPMG) method and extended the spin coherence time up to 17 $\mu s$
by TCPMG-2.\cite{key-8} 

For further increasing the spin coherence time for the thermometry,
in this work, we firstly studied the effects of the higher order TCPMG
method applied on the implanted single NV centers in diamond at room
temperature. In particular, a coherence time of 108 $\mu s$ was obtained
for TCPMG-8, about 14 times of the value 7.7 $\mu s$ for Thermal
Ramsey (T-Ramsey) method. This value corresponded to a thermal sensitivity
$\eta$ of 10.1 mK/Hz$^{1/2}$, which was comparable
with that of the native NV center in isotopically pure diamond.\cite{key-9,key-10}
Then we measured the temperature distribution on the surface of a
high-purity diamond in three different circumstances by performing
the TCPMG-3 pulse sequence measurement on the implanted NV center
array. The obtained thermal sensitivity $\eta$ reached 24 mK/Hz$^{1/2}$.
The results demonstrate that the TCPMG method can effectively extend
the spin coherence time of the implanted NV center, which paves the
way for using the implanted NV center in high-quality nanodiamonds\cite{key-29}
to practical temperature detection with nanoscale resolution and high-sensitivity. 

The ground state of the NV$^-$ is a spin
triplet ($S=1$), consisting three spin projection states $|m_{S}=0\rangle$
and $|m_{S}=\pm 1 \rangle$, which are split under spin-spin
interactions, exhibiting a zero-field splitting D=2$\pi \times$2.87 GHz at room temperature. The spin states can be spin polarized
and read out optically, and coherently controlled by microwave pulses.
The principle of temperature detection using NV center is based on
the temperature dependence of the zero-field splitting D, which depends
on the local lattice expansion induced by the temperature variation.\cite{key-5,key-6,key-7}
In particular, as has been reported, the value of D is linearly dependent
on the temperature with dD/dT = -74.2 kHz/K at the temperature from
280 K to 330 K.\cite{key-5} 

\begin{figure}
\includegraphics{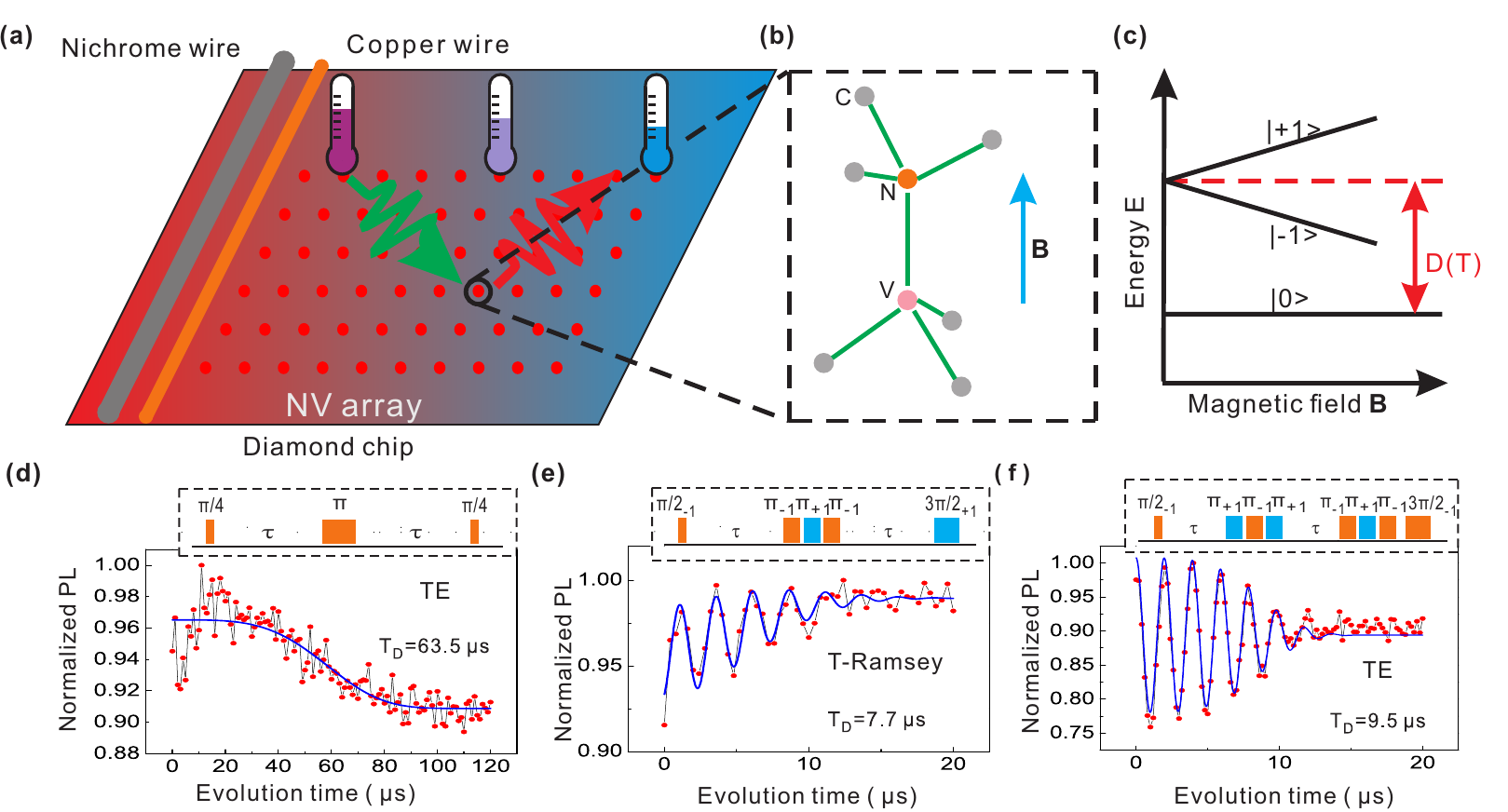}
\caption{The schematic of the NV center thermometry setup and the thermal pulse
sequence measurements. (a) The schematic of the NV center thermometry
setup. The implanted NV center array (red points) in a high-purity
diamond surface layer was used to sense local temperatures. (b) Atomic
structure of a nitrogen (N) - vacancy (V) center in diamond with an
arrow indicating the applied magnetic field B. (c) Ground state spin
energy levels are split in applied axial magnetic field Bz. (d) Thermal
echo measurement at zero magnetic field. The inset was the pulse sequence.
The coherence time for temperature sensing was 63.5 $\mu s$. (e)
Thermal Ramsey measurement in magnetic field. The inset was the pulse
sequence. The coherence time for temperature sensing was 7.7 $\mu s$.
(f) Thermal echo measurement in magnetic field. The inset was the
pulse sequence. The coherence time for temperature sensing was 9.5
$\mu s$. The blue lines were the fits to the data, the coherence
times were noted on the plots.}
\label{Figure 1}
\end{figure}

The scheme of the NV thermometry setup in our experiments was shown
in Figure 1 (a). The implanted NV center array in the high-purity
diamond was used to sense local temperature on the diamond surface
layer. The wavy green arrow represented the 532-nm optical excitation
laser which was used to polarize and read out the NV center spin states,
and the wavy red arrow represented the emitted fluorescence of the
NV center. A 20 $\mu m$ copper wire was placed on the diamond for
transmitting microwave to manipulate the spin states of the NV center
and a 40 $\mu m$ Nichrome wire placed beside the copper wire was
heated by a precision DC power source for sample temperature control.
An electromagnet generated a 32 G magnetic field for experiments in
magnetic field. 

The sample was a $2\times 2 \times 0.5 $ mm$^{3}$
(100) high-quality electronic grade diamond with natural isotopic
concentration of $^{13}$C (1.1\%) from Element Six ([N]
< 5 ppb). The NV center array was made by implanting 60 keV $^{14}N_{2}^{+}$
molecules with the fluence $2.25 \times 10^{11} $ $^{14}N_{2}^{+}/cm^{2}$ and the implantation angle $7^\circ$
through 45 nm diameter apertures patterned using electron beam lithography
in a 300-nm-thick polymethyl methacrylate (PMMA) layer deposited on
diamond surface.\cite{key-30} The average depth of the NV centers
was about 40 nm and the longitudinal and lateral straggling were about
11 and 9 nm, respectively, inferred from SRIM simulations. After implantation,
the sample was annealed at 1050 \textcelsius{} in a vacuum at $2 \times 10^{-5}$
Pa for 2 h to induce vacancy diffusion to form NV centers. Annealing
at this temperature could also reduce the total concentrations of
the paramagnetic residual defects to extend spin coherence times.\cite{key-31}
After oxidation at 430 \textcelsius{} in atmosphere for 2.5 h for
improving negatively charged NV centers conversion efficiency, the
sample was cleaned in a 1:1:1 boiling mixture of sulfuric, nitric,
and perchloric acid at 200 \textcelsius{} for one hour. The irradiation
dose used for NV center generation corresponded to about 5 nitrogen
atoms per aperture (45 nm diameter), so it was thought that the spin
bath for the NV center was mainly contributed by $^{13}C$ in the
diamond. In such a system, the Hamiltonian of NV center can be expressed
as \cite{key-8} 
\begin{equation}
H=D(T)S_{z}^{2}+g\mu_{B}\vec{B} \cdot \vec{S}+\vec{S}H_{B1}+H_{B2}
\label{eq:NV center Hamiltonian}
\end{equation}
where $\vec{S}$ is the NV center's electronic spin, g
= 2.00 is the electron g factor, $\mu_{B}$ is the Bohr magneton,
$\vec{B}$ is the applied magnetic field. The third term
describes hyperfine coupling of the NV center spin to the bath of
$^{13}C$ spins, and the last term describes the internal dynamics
of the $^{13}C$ nuclear spin bath. In general, the zero-field splitting
parameter D(T) depends on temperature T, axial electric field, and
strain. For temperature detection based on D(T), we resonantly manipulate
the spin states such that the unwanted relative phase are canceled,
getting the common phase factor $e^{-iDt}$, with
the phase proportional to D only. This detection produces a fluorescence
intensity ($I_{PL}$) oscillating between $I_{PL}$ ($m_{S}=0$) and
$I_{PL}$ ($m_{S}=\pm 1$) with the frequency given by $|D-\omega|$,
where $\omega$ is the microwave carrier frequencies used for spin
manipulation.\cite{key-8} When the change of oscillation frequency
is determined, the change of D and hence the corresponding local temperature
change can be deduced.

At zero magnetic field, by applying a thermal echo (TE) pulse sequence\cite{key-8,key-10},
as shown in Figure \ref{Figure 1}(d), the electronic spin of the
NV center was firstly initialized to a superposition state by a $\pi/4$
pulse. After half the total evolution time, a $\pi$ echo pulse
was used to reverse the population of the $|+1\rangle$ and $|-1\rangle$
states. After another half of the total free evolution time, the relative
phases between the $|\pm1\rangle$ levels, caused by quasi-static
fluctuations of magnetic field, were canceled, getting the common
phase factor $e^{-iDt}$, with the phase proportional
to D. The TE sequence produced a long coherence time of $\sim$ 63.5
$\mu s$, which was comparable with that of the native NV center in
high-quality diamond.\cite{key-8} However, there was not significant
oscillations when we detuned the microwave carrier frequencies ($\omega$)
from D. This result was confirmed by using other implanted single
NV centers in the sample. This was attributed to the implanted N electron
spin defects and the paramagnetic residual structure defects.\cite{key-8,key-10,key-32}
It was concluded that the TE method at zero magnetic field\cite{key-8,key-10}
was difficult to be used on the implanted NV center for temperature
sensing.

Then we applied the thermal Ramsey (T-Ramsey), TE, and TCPMG-N methods
to the implanted NV centers for temperature sensing under finite magnetic
fields. In these experiments, two microwave radiation fields with
different frequencies $\omega_{-1}$ and $\omega_{+1}$, in general,
are used to manipulate the transitions of $|0\rangle\leftrightarrow|-1\rangle$
and $|0\rangle\leftrightarrow|+1\rangle$, respectively. To induce
oscillations in $I_{PL}(t)$, both the microwave carrier frequencies
are slightly detuned from the corresponding resonance frequencies.
The $I_{PL}$ as a function of the free evolution time t follows the
equation\cite{key-8} 
\begin{equation}
I_{PL}=a\exp{(-(\frac{t}{T_{D}})^{n})} \cos{(2\pi ft+\varphi)}+b
\label{eq:fit function}
\end{equation}
where a, n, $\varphi$ and b are free parameters, $T_{D}$ is the
thermal pulse sequences coherence time. It is proved that the oscillations
frequency $f=|(\omega_{-1}+\omega_{+1})/2-D|$ (see Supporting Information).
So we can deduce the coherence time $T_{D}$ and oscillation frequency
$f$ (hence the zero-field splitting D) by fitting the recorded $I_{PL}(t)$
to Eq. 2. 

For the T-Ramsey experiment as shown in Figure 1(e)\cite{key-9},
the spin was firstly initialized into a superposition state $(|0\rangle+|-1\rangle)/\sqrt{2}$
by a $(\pi/2)_{-1}$ pulse. After half of the total evolution time,
a triple echo pulse sequence of the form $\pi_{-1}\pi_{+1}\pi_{-1}$
was applied to swap the population of the $|+1\rangle$ and $|-1\rangle$
states, where the $\pi_{-1}$ and $\pi_{+1}$ were the $\pi$-pulses
applied to manipulate the $|0\rangle \leftrightarrow |-1\rangle$ and
$|0\rangle \leftrightarrow |+1\rangle$ transition, respectively.\cite{key-8,key-9}
After another half the total evolution time, the relative phases between
the $|0\rangle$ and $|-1\rangle$ states were canceled, getting a
total phase factor $e^{-iDt}$, which was only related
to D, independent of low-frequency magnetic noise. We got a coherence
time of 7.7 $\mu s$ for the T-Ramsey sequence, which was larger than
the values of 1--5 $\mu s$ for nanodiamonds.\cite{key-9}
The reason was that, for nanodiamonds, the spin bath contains nuclear
spins ($^{13}C$), high concentration of electron spins ([N] about
100ppm), and surface layer spins.\cite{key-32,key-33}

\begin{figure}
\includegraphics{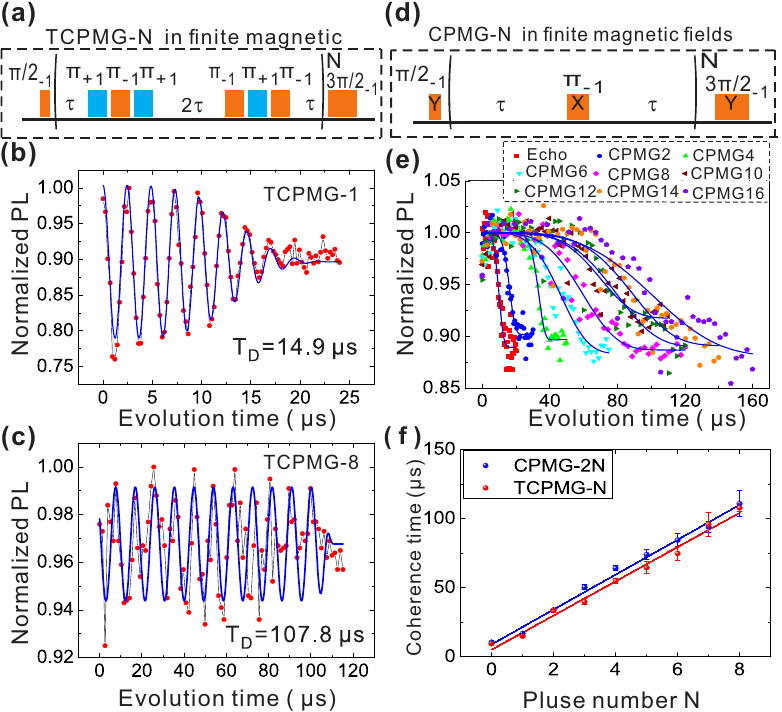}
\caption{TCPMG-N and CPMG-N pulse sequences measurements in magnetic field.
(a) The TCPMG-N pulse sequences. (b) The TCPMG-1 measurement. The
coherence time for temperature sensing was 14.9 $\mu s$. (c) The
TCPMG-8 measurement. The coherence time for temperature sensing was
107.8 $\mu s$. (d) The CPMG-N pulse sequences. (e) The coherence
decay curves of the Hahn echo and CPMG-N of a N from 2 to 16. The
coherence time of the CPMG-16 was 110.8 $\mu s$, which was ten times
longer than that for spin echo (10.5 $\mu s$). (f) The comparison
of the coherence time of the TCPMG-N and CPMG-2N for the same number
N. Both were linearly increasing with N. The blue and red points at
N=0 denoted the results for spin echo and TE, respectively.}
\label{Figure 2}
\end{figure}

For the TE (Figure \ref{Figure 1}(f)) and TCPMG-N experiments (Figure
\ref{Figure 2}(a)), the working principles are similar to that of
the T-Ramsey experiment .\cite{key-8} The key difference between
TE and TCPMG-N is that the TCPMG-N experiments invert the spin more
frequently (2N times) and hence more effectively eliminate higher
frequency magnetic noise, thus can extend the spin coherence time
for thermometry.\cite{key-8} By fitting the experimental resultsof the TE and TCPMG-N to the Eq. (2), the coherence times for thermometry
were derived. The coherence time of the TE measurement was 9.5 $\mu s$,
which was close to a value of 10.5 $\mu s$ obtained from the spin
echo experiment\cite{key-8} (Figure \ref{Figure 2}(e)). The TCPMG-1
(Figure \ref{Figure 2}(b)) and TCPMG-8 (Figure \ref{Figure 2}(c))
experiments extended the coherence time to 14.9 $\mu s$ and 107.8
$\mu s$, respectively. The coherence time of TCPMG-8 was about fifteen
times longer than that of the T-Ramsey. Utilizing the obtained experimental
data, the corresponding thermal sensitivity of the NV center $\eta$
can be derived from the following equation\cite{key-8} 
\begin{equation}
\eta=\sqrt{\frac{2(p_{0}+p_{1})}{(p_{0}-p_{1})^{2}}} \frac{1}{2\pi \frac{dD}{dT} \exp{(-(\frac{t}{T_{D}})^{n})} \sqrt{t}}\label{eq: thermal sensitivity}
\end{equation}
where $p_{0}$ and $p_{1}$ are the photon counts per measurement
shot for the bright and dark spin states, respectively. In the experiments,
we used the oil objective (NA = 1.4) and the obtained $p_{0}$ and
$p_{1}$ values were about 0.029 and 0.02, respectively. Thus derived
thermal sensitivity $\eta$ of the TCPMG-8 was 10.1 mK/Hz$^{1/2}$,
which was comparable with that of the native NV center in isotopically
pure diamond.\cite{key-9,key-10} 

Furthermore, the TCPMG method was compared with the conventional CPMG
method. In the CPMG-N experiments, as illustrated in Figure \ref{Figure 2}(d),
the microwave pulse phases of the beginning $(\pi_{Y}/2)_{-1}$ and
the final $(3\pi_{Y}/2)_{-1}$ pulses were Y phases, while the phases
of the echo pulses $\pi_{-1}$ were X phases.\cite{key-34} The coherence
times obtained in the CPMG-N experiments were increasing with the
$\pi$ pulse number N, and, in particular, the $T_{2}$ of the CPMG-16
was 110.8 $\mu s$, about ten times longer than the value 10.5 $\mu s$
for the Hahn echo. Considering the fact that there are two triple
$\pi$ pulses in a period of TCPMG, it would be more reasonable to
compare TCPMG-N with CPMG-2N. Figure \ref{Figure 2}(f) showed the
obtained dependence of the coherence times for these two methods with
the number N. It was found that in both cases the coherence times
increased almost linearly with N. This phenomenon of the CPMG was
similar to that for the native NV center in high-purity diamond,\cite{key-34}
but different from the $N^{2/3}$ dependence in low-purity diamond.\cite{key-35}
It was also noted that the coherence times of TCPMG-N were a little
shorter than that in the CPMG-2N, which might be caused by the pulse
imperfections, including pulse length, frequency imperfection, and
power fluctuation, since the TCPMG sequences were composed with more
microwave pulses. 

\begin{figure}
\includegraphics{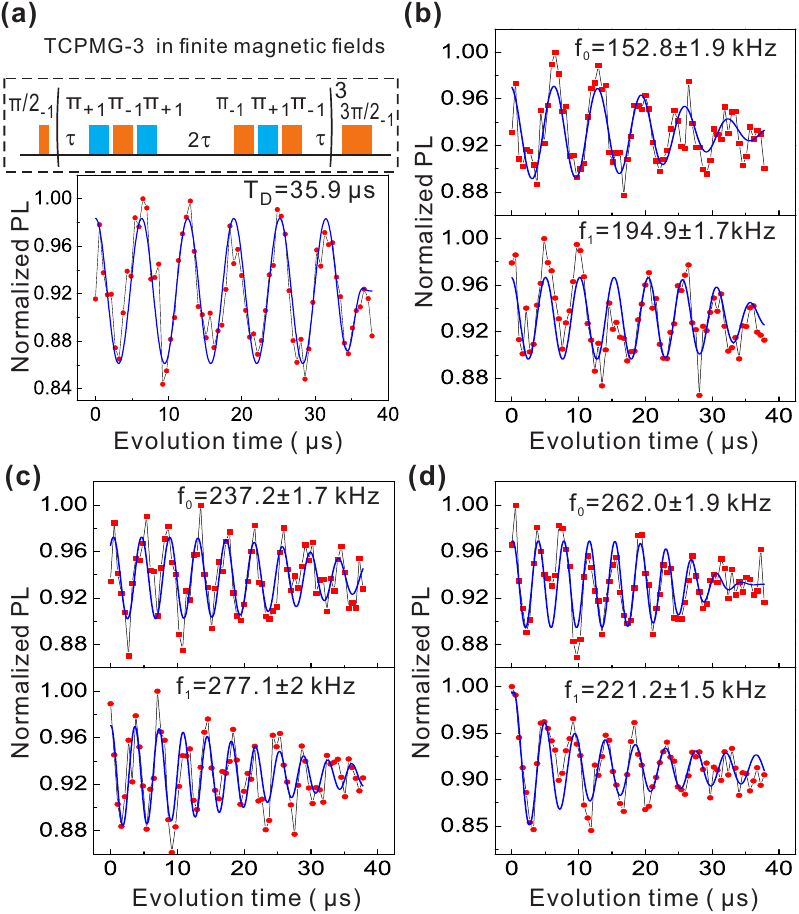}
\caption{Temperature detection based on the TCPMG-3 pulse sequence. (a) The
TCPMG-3 measurement in magnetic field. (b), (c) and (d) show the results
under three different microwave detunings, measured with the same
heating power on the Nichrome wire. (b) and (c) correspond to two
different positive detunings from D and (d) corresponds to a negative
detuning from D. The $f_{0}$ was the oscillation frequency in a case
without heating, and $f_{1}$ was the oscillation frequency with heating.
For all the TCPMG-3 experiments, each data point with $2\times 10^5$
averages. The measurement time per point was about 9 s, and total
time was about 10 minutes. }
\label{Figure 3}
\end{figure}

Then, we discussed, in more detail, the temperature detection through
measuring the changes in the $I_{PL}$ oscillation frequency by applying
TCPMG-3 on implanted NV centers. In Figure \ref{Figure 3}(a), the
upper part showed the pulse sequence of TCPMG-3 and the bottom part
showed the results of the TCPMG-3 measurement on a NV center. The
coherence time and the oscillation frequency, obtained from fitting
the results to Eq. (2), were 159.0 $\pm$ 1.0 kHz and 35.9 $\pm$
1.2 $\mu s$, respectively. In the experiments, we used the dry objective
(NA = 0.9) and the obtained $p_{0}$ and $p_{1}$ values were 0.022
and 0.017, respectively. The corresponding thermal sensitivity $\eta$
was derived from Eq.(3) to be 24 mK/Hz$^{1/2}$, which
was about 6 times improvement in comparison with that for the single
NV center nanodiamonds.\cite{key-9} Furthermore, the relation between
the change of the $I_{PL}$ oscillation frequency and the microwave
frequency detunings was examined. Figure 3(b,c,d) showed the results
for the sample both with and without heating, obtained under three
different detunings but the same heating condition. As shown in Figure
3(b) and 3(c), the changes of $I_{PL}$ oscillation frequency $f$
of the two different positive detunings (both of the microwave carrier
frequencies were larger than the corresponding resonance frequencies),
were 42.1$\pm$2.5 kHz and 39.9$\pm$2.6kHz, respectively,
while for the negative detuning (both of the microwave carrier frequencies
were less than the corresponding resonance frequencies), it was 40.8$\pm$2.4 kHz (see Figure 3(d)). It could be seen that the changes
of the oscillation frequency $f$ were nearly the same. This result
indicated that the change in oscillation frequency depends only on
the change in temperature, regardless of the microwave frequencies.
This was consistent with the theory described by Eq. (2). Using the
standard error derived from the fitting, we estimated that the temperature
precision was about 34 mK. 

\begin{figure}
\includegraphics{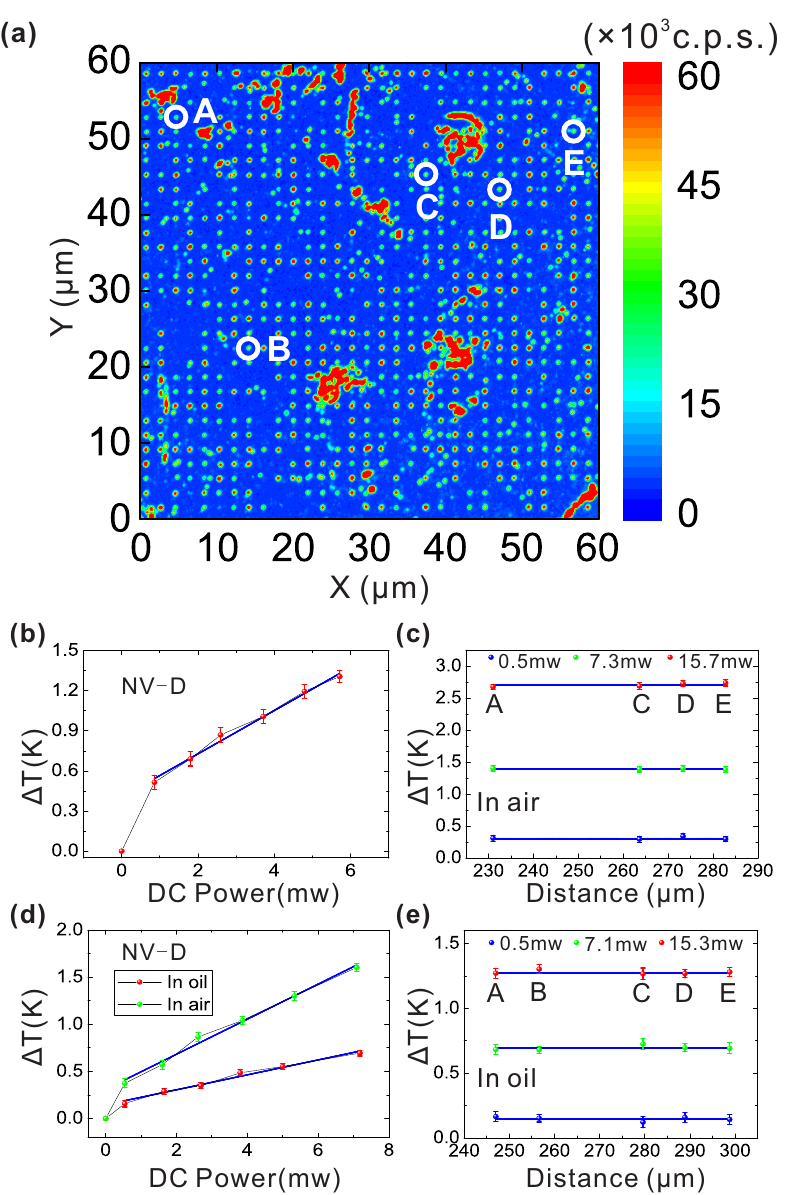}
\caption{Thermometry using the implanted single NV center array in the diamond
chip in air and in oil. (a) Confocal microscope fluorescence image
of the implanted NV center array in the high-purity diamond chip.
(b) The temperature detected by the center NV-D in the diamond chip
in air as a function of the heating power on the Nichrome wire. (c)
The temperatures measured by four single NV centers of the chip in
air as a function of the distance to the Nichrome wire under three
different heating powers. The blue lines for each heating power showed
the average temperature value of the four NV centers. (d) The temperature
detected by the NV-D as a function of the power on the Nichrome wire
for samples in air and oil. (e) The temperatures measured by five
single NV centers of the chip in oil as a function of the distance
to the Nichrome wire with three different powers. The blue lines for
each heating power showed the average temperature value of the five
NV centers.}
\label{Figure 4}
\end{figure}

To demonstrate detection of temperature distribution, the implanted
NV center array in the surface layer of a diamond chip was used, with
which the corresponding local temperatures were measured using TCPMG-3.
Figure \ref{Figure 4}(a) shows the confocal microscope fluorescence
image of an area of the sample with an implanted single NV center
array in its surface layer. The nearest separation of two NV centers
was 2 $\mu m$. The larger bright specks in the image were NV center
clusters formed during the implantation due to imperfect PMMA templet
layer deposited on the diamond chip. The sample was equipped with
a Nichrome wire heater, which was situated above the diamond surface
(Figure 1(a)) and arranged parallel to the Y axis. Five single NV
centers (named NV-A$\sim$E in Figure 4(a)) with their axes parallel
to the external magnetic field were selected to detect the local temperatures
in the diamond surface layer. For all the five NV centers, the coherence
time of the TCPMG-3 were about 35 $\mu s$. In the experiment the
sample was situated in air or in oil (Nikon microscope immersion oil).
Firstly, we detected the temperatures at four positions in the surface
layer of the diamond chip in air (Nichrome wire was placed 15 $\mu m$
above the surface of the diamond chip). Figure \ref{Figure 4} (b)
showed the local temperatures detected by the NV-D (273 $\mu m$ away
from the Nichrome wire) at various DC heating powers. It could be
seen that the temperature increased sharply at the beginning when
the heating power was low, then linearly increased as the DC heating
power increased. The sharp increase at the beginning might be due
to that the heat dissipation to the environment was less remarkable
as the temperature difference between the sample and the surroundings
was small at low heating power. On the other hand, as shown in Figure
\ref{Figure 4}(c), under each DC heating power, the temperatures
measured via four NV centers located at different distances from the
Nichrome wire were almost the same, even for an NV center 100 $\mu m$
away from the Nichrome wire. This result can be ascribed to the high
thermal conductivity $\kappa$ of the diamond (about 2000 W/mK). The
similar result was obtained by using the same NV-D (289 $\mu m$ away
from the Nichrome wire, Nichrome wire 23 $\mu m$ above the surface
of the diamond chip) with the diamond chip in oil (red points in Figure
\ref{Figure 4}(d)), except that the rate of the temperature variation
for sample in oil was less than that in air (green points). The temperatures
of the five single NV centers at different distances from the Nichrome
wire for the sample in oil (Figure \ref{Figure 4}(e)) were also almost
the same under the same DC powers, similar to the case of the sample
in air. 

\begin{figure}
\includegraphics{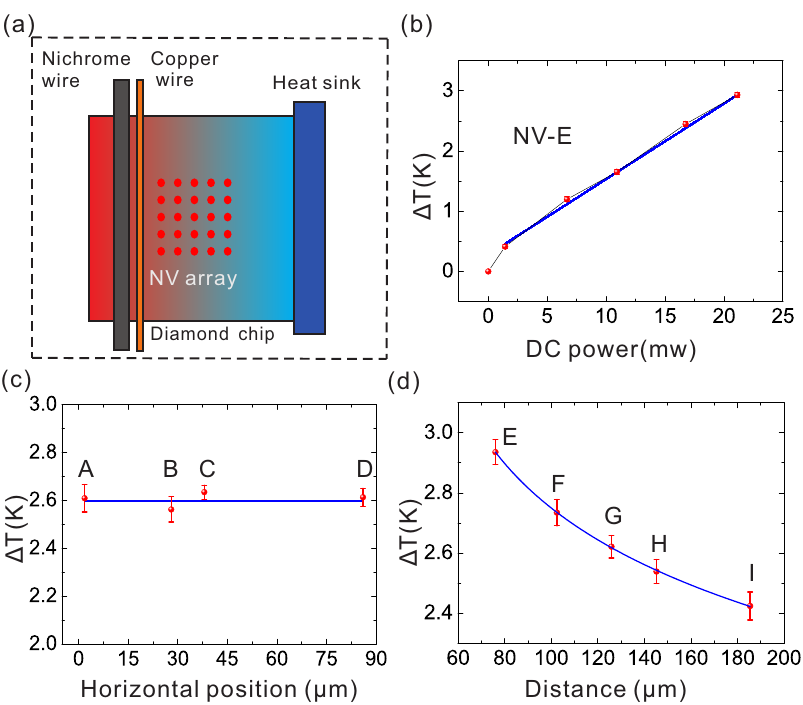}
\caption{Thermometry using implanted NV center array in the diamond chip, which
was located in air and connected to a heat sink. (a) The schematic
of the experimental setup for demonstrating the temperature distribution
detection. The setup is comprised of the diamond chip with an implanted
NV center array in it and, as shown in the figure, with a Nichrome
heating wire on the left side of the chip and a heat sink on the other
side. (b) The temperature detected by NV-E versus the heating power
on the Nichrome wire. (c) The temperature distribution along a direction
parallel to the Nichrome wire detected by four single NV centers (NV-A$\sim$D).
The blue line demonstrated that the temperature was nearly the same
along the direction parallel to the Nichrome wire. (d) The temperature
distribution along a direction perpendicular to the Nichrome wire
detected by five single NV centers (NV-E$\sim$I). The blue line was
the fitting of the experimental data.}
\label{Figure 5}
\end{figure}

Finally, we detected the temperature distribution on the surface layer
of the sample in air, but with a heat sink added on the other side
of the sample (Figure 5 (a)). The heat sink, a copper wire, was in
thermally contacted with a stable heat bath of a temperature 0 \textcelsius .
The Nichrome heating wire was 20 $\mu m$ above the surface of the
diamond chip. To detect the temperature distribution on the sample
surface, nine single NV centers (named NV-A$\sim$I, as shown in Supporting
Information Fig 3 and Fig 4) were selected to detect the temperatures
at the corresponding local positions. The axes of all the nine NV
centers were parallel to the magnetic field and the coherence times
measured with TCPMG-3 for these centers were about 35 $\mu s$. Figure
5(b) presented a typical DC power dependence of the temperature recorded
by NV-E (76 $\mu m$ away from the Nichrome wire), showing a relationship
similar to that for sample without the heat sink. Furthermore, Figure
\ref{Figure 5}(c) showed the temperatures measured by using four
single NV centers, NV-A$\sim$D, that were located at about the same
distance (about 96 $\mu m$) away from the Nichrome wire from the
Nichrome wire with a DC heating power of 20.66 mw. Obviously the temperatures
were almost the same, which was in accordance with the geometry of
these NV centers: the same distance from the heating wire. However,
as shown in Figure 5(d), the temperatures varied along the perpendicular
direction, as represented by temperatures of the five single NV centers,
NV-E$\sim$I, at different distances to the Nichrome wire that was
heated with a DC power of 21.15 mw. According to the steady-state
heat conduction equation, the temperature profile on the diamond surface
layer follows the expression $\Delta T=a \frac{Q}{\kappa} \ln{r} +b$,
where a, b are free parameters, $Q$ is the heat flux, and $\kappa$
is the thermal conductivity of diamond and r is the distance between
the NV center and the Nichrome wire. It can be seen from the figure
that the experimental data was fitted very well with the expression.
The experiments showed the effectiveness of the TCPMG method for high-sensitivity
temperature detection when performed on the implanted NV centers in
high-purity diamond. 

In summary, we studied thermometry based on the implanted single nitrogen
vacancy (NV) center in diamond by using the TCPMG method in a static
magnetic field. It was demonstrated that the spin coherence time for
thermometry was extended up to 108 $\mu s$ for TCPMG-8, which was
around 14 times of the value for T-Ramsey method (7.7 $\mu s$). This
value corresponds to a thermal sensitivity 10.1 mK/Hz$^{1/2}$,
which was comparable with that for the isotopically pure diamond.\cite{key-9,key-10}
We measured the temperature distributions on the diamond chip surface
in three different circumstances using the TCPMG-3 pulse sequence
on the implanted NV center array. The achieved thermal sensitivity
was 24 mK/Hz$^{1/2}$, which was about 6 times improvement
in comparison with that for the single NV center nanodiamonds.\cite{key-9}
The experiment implies the feasibility for using implanted NV centers
in high-quality diamonds to detect temperatures with high-sensitivity.

It is expectable that using higher order TCPMG, isotopically pure
diamond and technique of higher photon collection efficiency, such
as solid immersion lenses,\cite{key-36} the thermal sensitivity can
be further improved to submK/Hz$^{1/2}$. Combining the
TCPMG method and the implanted NV center in high-purity nanodiamonds,\cite{key-29}
high performance temperature sensors with higher precision, nanoscale
spatial resolution, outstanding sensor photostability and chemical
inertness can be constructed, which can be applied to nanoscale temperature
detection in a wide variety of systems, including biology,\cite{key-10}
chemistry, material science and microelectronics systems. The TCPMG
thermometry could also be applied to other solid-state quantum spin
systems such as point defects in silicon carbide\cite{key-11,key-37}
for temperature sensing.

\section*{ACKNOWLEDGMENTS}
We thank Qi Zhang, Fazhan Shi, Pengfei Wang, Jinming Cui, Zhaojun
Gong and Jie You for their help in building the experiment setup and making the sample. We also thank
Lei Zhou for his help in the preparation of this paper. This work
was supported by the National Basic Research Program of China (2013CB921800,
2011CB921400) and the Natural Science Foundation of China (Grant Nos.
11374280, 50772110).

\end{document}